\documentstyle[11pt,newpasp,twoside,epsf]{article}

\def\hkpc{$~h_{70}^{-1}$ kpc}
\def\apm{APM08279+5255}

\def\edcomment#1{\iffalse\marginpar{\raggedright\sl#1\/}\else\relax\fi}
\reversemarginpar

\begin{document}
\title{The Structure of High Redshift Galactic Halos}
\author{Sara L. Ellison}
\affil{P. Universidad Catolica de Chile, Casilla 306, Santiago 22, Chile;\\
European Southern Observatory, Casilla 19001, Santiago 19, Chile;\\ 
University of Victoria, 3800 Finnerty Rd., Victoria, BC, V8P 1A1, Canada.}
\author{Rodrigo Ibata}
\affil{Observatoire de Strasbourg, 11, rue de l'Universit\'e, F-67000, 
        Strasbourg, France}
\author{Max Pettini}
\affil{Institute of Astronomy, Madingley Rd, Cambridge, CB3 0HA, U.K.}
\author{Geraint F. Lewis}
\affil{Institute of Astronomy, School of Physics, A~29, University of Sydney, NSW 2006, Australia}
\author{Bastien Aracil, Patrick Petitjean}
\affil{Institut d'Astrophysique de Paris -- CNRS, 98bis Boulevard 
        Arago, F-75014 Paris, France}
\author{R. Srianand}
\affil{IUCAA, Post Bag 4, Ganeshkhind, Pune 411 007, India}

\begin{abstract}

Observations of multiple or lensed QSOs at high redshift can be used to
probe transverse structure in intervening absorption systems.  Here
we present results obtained from STIS spectroscopy of the unique triply
imaged QSO \apm\ at $z_{\rm em} = 3.9$ and study the coherence scales of
intervening low and high ionization absorbers.

\end{abstract}

\section{Introduction}

\begin{figure}[t]
\plotfiddle{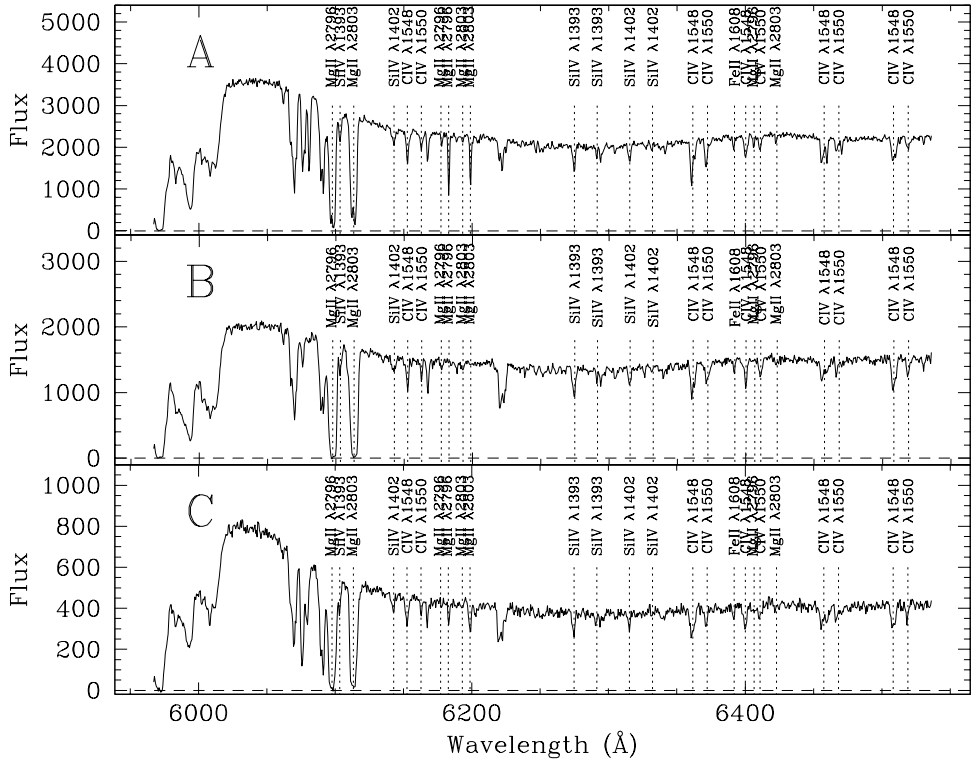}{10.0cm}{270}{55}{55}{-220}{300}
\caption{Example of the extracted STIS spectra for the 3 components of
the triply imaged QSO with intervening absorbers labelled.}
\end{figure}

As we have heard throughout this meeting, absorption line spectroscopy
is a powerful and accurate technique for determining abundances in the
interstellar and intergalactic media.  The advantages of this technique
become even more pronounced for applications in the high redshift universe
where the familiar UV transitions used to determine gas phase
abundances are shifted into the optical.  In this regime these transitions 
can be studied with efficient CCD detectors mounted on large ground-based 
telescopes.  By exploiting these advantages, enormous progress has been in the
last decade in determining the metal abundances of high redshift
absorption systems with accuracies that rival those possible for Galactic
stars.  Given the impressive bulk of abundance data for a range of
chemical elements, it is somewhat ironic that other (perhaps
more fundamental) properties of absorption galaxies such as size,
luminosity and
morphology remain relatively unconstrained, with information available
for only a handful of low redshift systems.

One technique that has been used to expand upon the single dimension
studies of QSO absorbers is to use multiple, but closely spaced 
lines of sight, either
via gravitational lensing or chance alignments of quasars.
The multiple lines of sight (LOS) will then intersect intervening
galaxies at different transverse locations, offering the opportunity
to probe the absorber on parsec to kiloparsec scales (e.g. Rauch 2002).
In order to improve our knowledge of the sizes and internal velocity
structure of high redshift absorbing galaxies,
we have undertaken a program with HST to obtain spatially
resolved spectra of the lensed QSO \apm\ ($z_{\rm em}=3.9$), 
a unique triply imaged QSO 
(Ibata et al. 1999; Lewis et al. 2002) with image separations ranging
between 0.15 and 0.38 arcsec.  This sightline is particularly rich with 
absorption
systems over a large redshift range (Ellison et al. 1999a,b) and 
therefore offers an exceptional
opportunity to study the intervening gas on a range of transverse scales.

We assume the `concensus' cosmology of $\Omega_M = 0.3$, 
$\Omega_{\Lambda}=0.7$, $H_0 = 70$\,km~s$^{-1}$~Mpc$^{-1}$ throughout
and convert literature values to this scale when making comparisons.
We also assume that the redshift of the lensing galaxy is $z_{\rm lens} 
= 1.062$.

\section{ An Overview of the Data}

We have obtained spatially resolved spectra of the 3 components
of \apm\ with STIS on
board the HST.  A total of 25 orbits
was divided between 5 wavelength settings to achieve a total 
coverage of $\sim 6000$ to $\sim 8600$\,\AA.  The spectral resolution 
is 1.6\,\AA\ 
FWHM and the typical S/N per pixel ranges from 60 to 20.
In Figure 1 we plot an example of the extracted STIS spectra with
absorption systems labelled.
Full details of the data acquisition and reduction, as well as measurements
of absorption line equivalent widths (EWs) can be found in Ellison 
et al. (2003).

\section{Comparison of Low and High Ionization Absorbers}

\begin{figure}
\plotfiddle{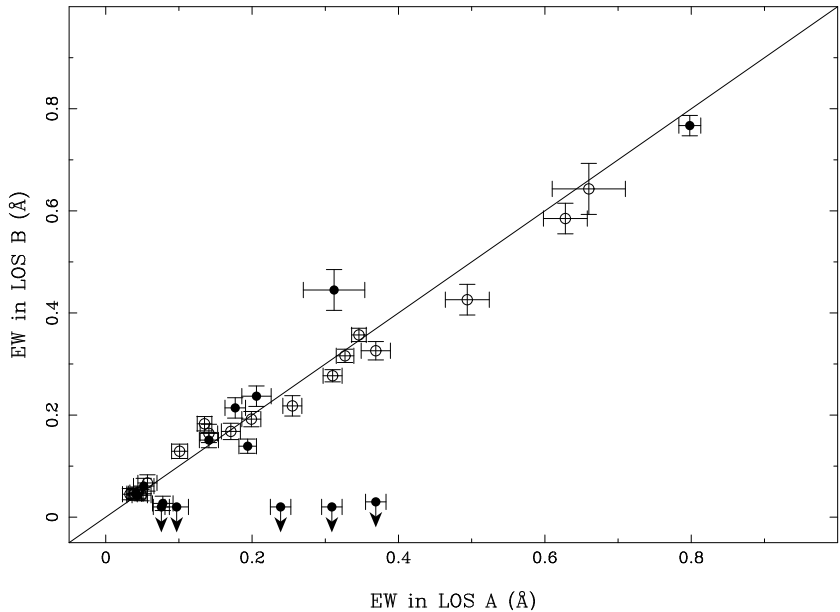}{7.5cm}{270}{45}{45}{-180}{260}
\caption{Rest frame EW comparison for absorption systems in two of the
components (A and B) of the lensed QSO \apm. 
High ionization systems are plotted with open points, low ionization
systems are filled.}
\end{figure}

In Figure 2 we show a comparison of EWs for low (Mg~II and Fe~II)
and high (C~IV and Si~IV) ionization absorbers in 2 of the 3 LOS, 
labelled A and B (see Ibata et al. 1999 for NICMOS imaging of the
three closely spaced components).
The striking result of this plot is that the high ionization systems
trace each other relatively closely in terms of line strength, whereas
there can be marked differences in the low ionization systems.  
However,
one should bear in mind that the redshifts, and therefore transverse
separations, are different for these systems; C~IV absorbers are observed
with typical separations of several hundred pc, whereas Mg~II systems
probe scales of a few kpc.  Nonetheless, these results confirm earlier
work by Rauch et al. (1999, 2000, 2001) that cooler, low ionization
gas is clumped on smaller scales than the warmer, more highly ionized
component.

\section{Coherence Scales of Absorbers}

\begin{figure}
\plotfiddle{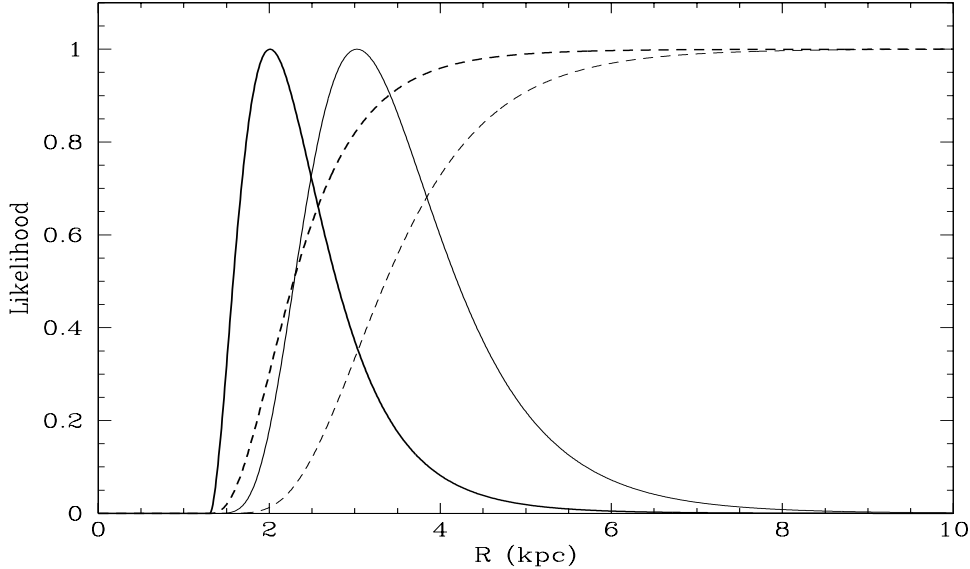}{7.0cm}{270}{50}{40}{-200}{230}
\caption{
Maximum likelihood distribution (thin solid line)
and cumulative distribution (dashed line) for spherical
halos normalized to the peak value. The most likely coherence
radius is found
to be R=3.0 \hkpc\ with 95 \% confidence limits
of 2.1 and 6.2 \hkpc.  The
bold lines show the results obtained if we restrict ourselves to 
weak (EW$<$0.3 \AA) Mg~II systems only, for which we deduce a
most likely scale of $R$=2.0 \hkpc\ with
95\% confidence limits  of 1.5 and 4.4 \hkpc.  }
\end{figure}

\begin{figure}
\plotfiddle{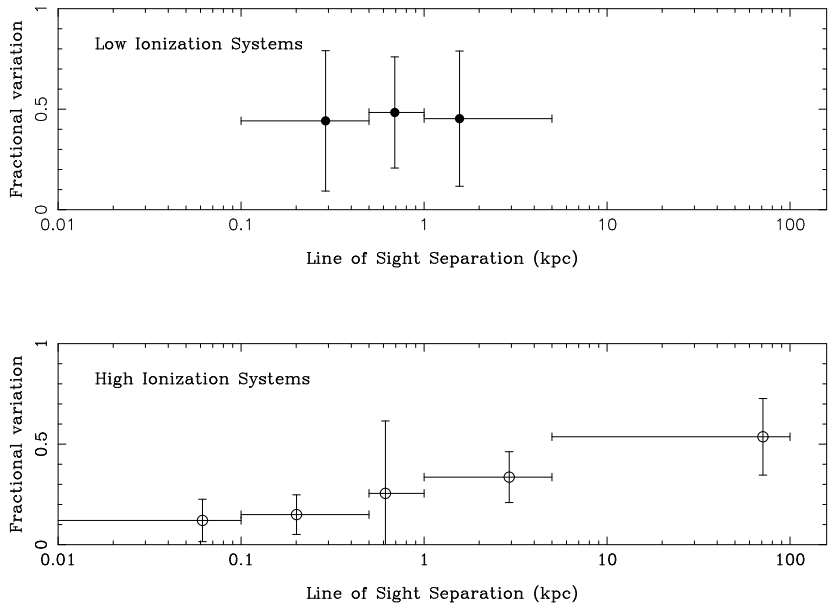}{7.5cm}{270}{45}{45}{-180}{260}
\caption{  Fractional EW 
variations for low and high ionization
systems from this work and the literature (Lopez et al. 2000;
Rauch et al. 2001; Rauch et al. 2002; Churchill et al. 2003).}
\end{figure}

C~IV absorbers have been studied in lensed sightlines on a range of scales
that confirm coherence over $\sim$ 100\,\hkpc\ (e.g. Petitjean 
et al. 1998; Lopez et al. 2000).  The transverse separations
towards \apm\ are considerably smaller than this, and, as is evident from
Figure 2, a high degree of coherence is seen at $z\sim3$ on scales of
100s of pc.  However, little work has been done previously 
on the low ionization systems due
to a paucity of statistics.  For the first time, we have a sufficient
number of Mg~II systems to attempt a maximum likelihood estimate of
coherence scale based on the number of coincidences and anti-coincidences
in pairs of LOS using the technique of McGill (1990).  For the full
sample of Mg~II absorbers we find a most likely coherence radius of  
$R = 3.0\,h_{70}^{-1}$\,kpc, with 95\% confidence limits of 2.1 and 
6.2\,kpc (Figure 3).  This is apparently at odds with the size of
Mg~II halos, $\sim$50 kpc, inferred by Steidel, Dickinson, \& Persson 
(1994) from the number density of absorbers and galaxy impact parameters.
However, we can divide our sample of Mg~II absorbers into `strong'
(EW $\ge 0.3$ \AA, the typical EW of most absorption line system
surveys) and weak (EW $< 0.3$ \AA) subsets.  We can only place
a lower limit of $R > 1.4\,h_{70}^{-1}$\,kpc on the strong absorbers
because all systems are seen in all LOS.  For weak systems, we infer
a most likely coherence radius of  $2.0\,h_{70}^{-1}$\,kpc with 95\% 
confidence limits of 1.5 and 4.4\,kpc (see Figure 3).  This
could be explained if the filling factor of the low ionization gas
decreases towards the outer part of the galaxy, leading to a smaller
coherence scale for low EW absorbers.  This scenario seems to be
qualitatively supported by the observations of Churchill et al. (2000)
that impact parameter is anti-correlated with $N$(Mg~II).

We can also examine whether absorption systems exhibit larger transverse
variations as a function of LOS separation. In Figure 4 
we have plotted the mean fractional variation in EW for absorption systems
in this study and for data compiled from the literature as a function of
transverse separation.  It can be seen 
that for the low ionization systems the mean variation is the same for all
separations within the large scatter.  This is probably due to the fact that
individual components cannot be traced over more than a few hundred
pc (Rauch et al. 2002).  The high ionization systems, however, show a 
clear trend for larger variations with increasing LOS separation.  

These results are qualitatively similar to observations of the Galaxy,
where cold gas traced by Na~I shows fine spatial structure on AU scales
(e.g. Lauroesch, Meyer \& Blades 2000).  Slightly warmer, but still 
predominantly neutral, gas
marked by Ca~II (which is comparable to the phase traced by Mg~II and
Fe~II studied here) also shows small scale variations, but is less
clumpy than the cold gas (e.g. Redfield \& Linsky 2002).  
The warm ionized component, traced for example
by C~IV, is much more smoothly distributed and shows far less variation
(Savage, Sembach \& Lu 1997).

These results show the power of multiple LOS
spectroscopy for providing information on the spatial extent and
kinematics of intervening galaxies which cannot be obtained
by other means.

\acknowledgments{SLE is very grateful to the IAU for a generous travel
grant and to the SOC and LOC for their help in facilitating attendance
at this enjoyable symposium.}


\begin{references}
\reference Churchill, C. W., Mellon, R. R., Charlton, J. C., Jannuzi, B. T.,
Kirhakos, S., Steidel, C. C., Schneider, D. P.,  2000, \apj, 543, 577
\reference Churchill, C. W., Mellon, R. R., Charlton, J. C., Vogt, S.,
 2003, \apj, 593, 203
\reference Ellison, S. L., Ibata, R., Pettini, M., Lewis, G., Aracil, B.,
Petitjean, P., Srianand, R., 2003, \aap, submitted.
\reference Ellison, S.\ L., Lewis, G.\ F., Pettini, M., Chaffee, F.\ H., 
Irwin, M.\ J., 1999a, \apj,  520, 456
\reference Ellison, S.\ L., Lewis, G.\ F., Pettini, M., Sargent, W.\ L.\ W., 
Chaffee, F.\ H., Foltz, C.\ B., Rauch, M., Irwin M.\ J., 1999b, PASP,  111, 946
\reference Ibata, R., Lewis, G. F., Irwin, M. J., Lehar, J., Totten, E. J.,
1999, \aj, 118, 1922
\reference Lauroesch, J. T., Meyer, D. M., \& Blades, J. C., 2000, ApJ,
543, L43
\reference Lewis, G. F., Ibata, R., A., Ellison, S. L., Aracil, B., 
Petitjean, P., Pettini, M., Srianand, R., 2002, \mnras, 334, L7
\reference Lopez, S., Hagen, H.-J., Reimers, D., 2000, \aap, 357, 37
\reference McGill, C., 1990, \mnras, 242, 544
\reference Petitjean, P., Surdej, J., Smette, A., Shaver, P., M\"{u}cket, J.,
Remy, M., 1998, \aap, 334, L45
\reference Rauch, M., 2002, ASP Conference Proceedings, `Extragalactic Gas
at Low Redshift', Eds, J. Mulchaey, J. Stocke, 254, 140
\reference Rauch, M., Sargent, W. L. W., \& Barlow, T. A., 1999, \apj, 515, 500
\reference Rauch, M., Sargent, W. L. W., \& Barlow, T. A., 2001, \apj, 554, 823
\reference Rauch, M., Sargent, W. L. W., Barlow, T. A., Simcoe, R. A., 
2002, \apj, 576, 45
\reference Redfield, S. \& Linsky, J. L., 2002, ApJS, 139, 439
\reference Savage, B. D., Sembach, K. R., \& Lu, L., 1997, AJ, 113, 6 
\reference Steidel, C. C., Dickinson, M., \& Persson, E. 1994, \apj, 437, L35
\end{references}
\end{document}